\documentclass[pra,aps,twocolumn,amsmath,amssymb]{revtex4}

\usepackage[dvips]{graphicx}% Include figure files
\usepackage{dcolumn}% Align table columns on decimal point
\usepackage{bm}% bold math
\usepackage[extension=xxx]{hyperref}

\newcommand{\beq}{\begin{equation}}
\newcommand{\eeq}{\end{equation}}

\newcommand{\beqa}{\begin{eqnarray}}
\newcommand{\eeqa}{\end{eqnarray}}

\def\beq{\begin{equation}}

\usepackage{color}
\begin{document}

\title{Direct tunneling delay time measurement in an optical lattice}
\date{\today}

\author{A. Fortun, C. Cabrera-Guti\'errez, G. Condon, E. Michon, J. Billy and D. Gu\'ery-Odelin}

\affiliation{Laboratoire Collisions, Agr\'egats, R\'eactivit\'e, IRSAMC, Universit\'e de Toulouse, CNRS, UPS, France}

\begin{abstract}
We report on the measurement of the time required for a wave packet to tunnel through the potential barriers of an optical lattice. 
The experiment is carried out by loading adiabatically a Bose-Einstein condensate into a 1D optical lattice. A sudden displacement of the lattice by a few tens of nm excites the micromotion of the dipole mode. %The frequency of this oscillatory collective mode is used to calibrate in-situ the optical lattice depth.
 We then directly observe in momentum space the splitting of the wave packet at the turning points and measure the delay between the reflected and the tunneled packets for various initial displacements. Using this atomic beam splitter twice, we realize a chain of coherent micron-size Mach-Zehnder interferometers at the exit of which we get essentially a wave packet with a negative momentum, a result opposite to the prediction of classical physics.
\end{abstract}


\maketitle

The question of the time required for a particle to tunnel through a barrier was addressed in the early days of quantum mechanics \cite{paper32}. The first quantitative study providing a finite tunneling time dates back to 1962 \cite{hartman}. The difficulty to define such a time lies in the fact that time is not in general associated with an ordinary observable in quantum mechanics \cite{controversial}.  As a result there is not a unique definition of tunneling time (see discussion in \cite{GonBook}).

Experimentally, this question was addressed in condensed matter, attosecond (as) physics and optics. In condensed matter physics, it has been investigated when performing time measurements in electronic systems on sub-ns timescale. Such studies have been carried out for instance in GaAs/GaAlAs hetereostructures \cite{gueret} and in biased Josephson junctions \cite{esteve}. 

More recently, attosecond tunneling spectroscopy shed light on how electrons quit their atomic binding potential under the action of a strong optical field. Such experiments constitute a real-time observation of light-induced electron tunneling on a timescale of a few hundreds of as \cite{atto,atto2}, and even a few tens of as with angular streaking techniques \cite{landsman}.

The tunnel problem in quantum mechanics can also be mapped on a wide number of situations in electromagnetism including frustrated total internal reflection \cite{golub,balcou}, transmission in 1D photonic bandgaps \cite{Steinberg,phbandgap} or transmission through a waveguide beyond the cutoff \cite{ranfagni}. The corresponding timescales are ns for microwaves and fs in the visible range. These experiments brought to the fore the question of superluminal motions. 

The tunnel effect plays a key role in the dynamics of degenerate gases trapped in optical lattices \cite{bloch2,morsch}. The tunneling rate $J$ can be directly measured in situ \cite{arimondo1} and engineered by an appropriate shaking of the periodic potential to design effective Hamiltonians and produce synthetic Gauge Fields \cite{arimondo2,sengstok,dalibard}. The interplay between tunneling rate and interactions is at the heart of the many body physics investigated with cold atoms in optical lattices \cite{bloch1,Josephson,nagerl,bloch3}, including in the presence of dissipation \cite{ott2013}. However, to date, no direct tunneling time measurement has been performed in this domain.

\begin{figure}[h!]
\centering
\includegraphics[width=7cm]{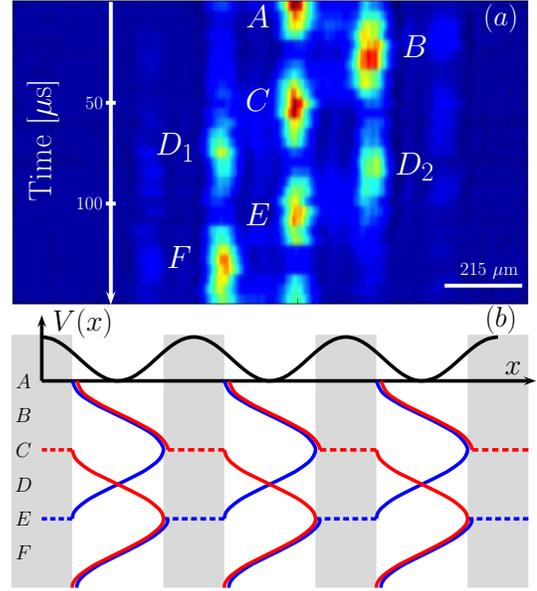}
\caption{(a) Time sequence showing the evolution of the wave packet in the lattice after a sudden displacement  (5 $\mu$s time interval between each picture). The data presented are averaged over 2 iterations. The images are taken after a 25 ms time-of-flight. (b) Sketch of the center-of-mass motion of the packets in each well. Their splitting by tunneling at the turning points is represented as dashed lines.}
\label{fig1}
\end{figure} 

In this article, we propose a direct measurement of tunneling time of massive particles by studying the out-of-equilibrium dynamics of a chain of coupled Bose-Einstein condensates (BECs) in an optical lattice. In contrast with the methods explored in other fields, we choose parameters so that roughly half the wave packet tunnels through the barrier. Using both the quasi-isochronism of oscillations in the lattice and the packet that has not tunneled as a reference, we infer precisely the duration of the tunneling process which could be up to 27 $\mu$s, i.e. 1/4 of the oscillation period.
 
Our experiment starts with a rubidium-87 BEC produced in a hybrid trap \cite{spielman2009}. In short, $2 \times 10^9$ atoms from a magneto-optical trap are loaded into a magnetic quadrupole whose gradient is ramped up to 1.8 T/m. Microwave evaporation is then performed over 15 s to decrease the temperature from 300 $\mu$K to 30 $\mu$K. Atoms are subsequently transferred to a crossed dipole trap, formed by two 1064 nm laser beams, which are respectively aligned along an horizontal axis and at $35^o$ from the vertical, while maintaining a magnetic gradient that approximately compensates for the gravity. The evaporation in this latter trap yields a pure BEC of $10^5$ atoms in the low field seeker state $F=1,m_F=-1$.
 
We then superimpose to the horizontal guide a 1D optical lattice made of two counter-propagating laser beams at 1064 nm (lattice spacing $d=532$ nm). The lattice is switched on adiabatically in 11 ms using a S-shape time variation of the intensity of the beams \cite{sshape}. The relative phase, $2\theta$, of the two laser beams is controlled using two phase-locked synthesizers that drive the acousto-optic modulators placed on each beam before they enter the vacuum cell. The 1D potential experienced by the atoms reads
\begin{equation}
V(x)=\frac{1}{2}m\omega_{\rm ext}^2x^2-sE_L \cos^2\left(  \frac{\pi x}{d} + \theta(t)\right)
\end{equation}
where the first term accounts for the weak external potential in which the BEC is produced. The external angular frequency $\omega_{\rm ext}$ results from the combined trapping of the dipole trap and the magnetic gradient ($\omega_{\rm ext}=2\pi\times 25$ Hz). The second term accounts for the lattice potential with a tunable time-dependent phase $\theta(t)$. $E_L=h^2/(2md^2)$ is the lattice characteristic energy and $s$ a dimensionless parameter characterizing the lattice depth. 

The lattice can be suddenly spatially shifted by a phase $\theta(0^+)=\theta_0$ ($\theta_0=90^o$ corresponds to a shift of half a period). This phase jump is performed on an ultrashort timescale (a few ns) compared to all other timescales in the system. This shift triggers an oscillation of the BEC in the lattice. We analyze this oscillation by suddenly switching off all trapping potentials followed by a $t_{TOF}=25$ ms time-of-flight. This provides a measurement of the in-trap atomic momentum distribution. The observed pattern results from an interference of the BECs located at each lattice site and is characterized by peaks associated to the momenta $p_n=nh/d$ with $n$ integer (positive or negative) and separated by a distance $ht_{TOF}/(md)=215$ $\mu$m \cite{inguscio2001a}.

For the experimental data presented in Fig.~\ref{fig1}, we used an initial shift of $d/4$ (i.e. $\theta_0=45^o$) and repeated the sequence for various holding times in the shifted lattice. In this way, we access to the dynamics of the BEC inside the lattice.  After $\sim 25$ $\mu$s of evolution in the lattice, the atoms return to the bottom of the potential wells, acquiring their maximum momentum in the meantime. This results in a shift of the maximum peak towards the right (see point $B$ in Fig.~\ref{fig1}). Atoms then evolve towards the opposite turning point that is reached at $C$. A quarter of period later, we observe two peaks ($D_1$ and $D_2$) and not just one, revealing in momentum space the splitting of the wave packet that occurred at $C$: a part of the wave packet ($D_1$) continued its oscillatory motion and another part of the wave packet ($D_2$) tunneled through the barrier and kept as a consequence its positive momentum. The potential barrier therefore acts as a beam splitter. 
Interestingly, the second interaction with the beam splitter that occurs at the next turning point ($E$) generates, by constructive interference, only one packet with a negative momentum.

%This important prerequisite enables a quantitative study of the properties of the tunnel effect observed in our experiment. We will then interpret the interference effect that is observed at the end of the sequence.
%In particular, it allows for a direct measurement of the tunneling delay time between the tunneled packet that has tunneled and the one that has not. 

\begin{figure}[h!]
\centering
\includegraphics[width=8cm]{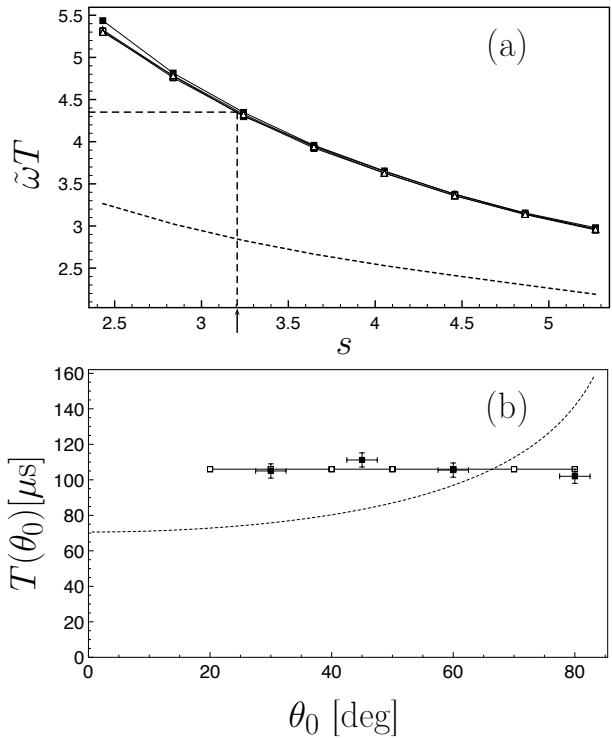}
\caption{(a) Collective dipole mode period measured in dimensionless units as a function of the lattice depth (normalized to the lattice characteristic energy $E_L$) for various values of the external potential and the interactions: $\tilde \omega _{\rm ext}=1/50$ (triangle), $\tilde \omega _{\rm ext}=1/262$ (square), $\beta=0.1$ (open symbol) and $\beta=1$ (filled symbol). The proximity of the curves enables one to extract precisely the optical lattice depth (dashed line). The dotted curve corresponds to the classical oscillation period \cite{classical}. (b) Oscillation period as a function of the initial angle $\theta_0$ (which accounts for the initial sudden displacement of the lattice): experimental points (filled square), numerical simulation with the calibration presented in (a) (empty square) and classical prediction (dotted line).}
\label{fig2}
\end{figure}

We start by investigating the oscillatory motion inside the lattice. The dipole motion of a condensate in an optical lattice has been studied experimentally in Refs.~\cite{inguscio2001b,science2001,porto2005} for an initial displacement of the condensate by a few tens of micrometers, i.e. more than its size, and in the low depth limit $s<1.2$ (with our notation). To compare the oscillations that result from such a large displacement, a hydrodynamic formalism was used, which substitutes a continuum formulation to the discretized formulation involving the wave functions in each site \cite{stringari2002,Clark2006}. The frequency of the dipole mode in this model is given by $\omega_{\rm dip}=(m/m^*)^{1/2}\omega_{\rm ext}$ where $m^*$ is the effective mass, a quantity that was extracted from the experimental data \cite{inguscio2001b}. In such experiments, the micromotion of the wave packets inside each lattice site does not play any role on the center-of-mass motion. This is to be contrasted with the case studied in this article for which we have a small displacement (a few tens of nm) and a deeper optical potential ($s\sim 3.2$). Under those conditions, the micromotion plays a major role and the continuous approach of Ref.~\cite{stringari2002} is no more adapted. Furthermore, the quantum dynamics that takes place inside each well cannot be mapped on the classical dynamics \cite{classical}. Indeed, despite the relatively large lattice depth, the tunnel effect still plays an important role on the out-of-equilibrium dynamics that we investigate. For an initial shift of $d/4$, it increases the oscillation period compared to the classical case (see Fig.~\ref{fig2}).

We investigated quantitatively the extra inertia provided by the lattice using numerical simulations of the 1D Gross-Pitaevskii equation \cite{GPELab}. For this purpose, we solve the dimensionless equation: $i\partial_{\tilde t} \psi = [(-\Delta + \tilde \omega^2_{\rm ext}X^2)/2 - \gamma\cos^2(\pi X/4+\varphi) + \beta |\psi|^2]\psi$, where the time is normalized to $\tilde t = \tilde \omega t$ with $\tilde \omega^{-1} = md^2/(16 \hbar)=24.3$  $\mu$s for our parameters, $ \tilde \omega_{\rm ext}=\omega_{\rm ext}/\tilde \omega$, and $\gamma=\pi^2s/8$. The precise value of the interaction strength $\beta$ requires a model for the effective reduction of dimension \cite{Pethick} and a precise knowledge of the number of atoms per site which differs from site to site because of the external confinement. Assuming a transverse Thomas-Fermi profile, we find $\beta \alt 1$. The dipole oscillation period $T$ normalized to $\tilde \omega^{-1}$ is plotted in Fig.~\ref{fig2}a as a function of the depth parameter $s$ for various values of the interaction parameter $\beta$ and external confinement $\tilde\omega_{\rm ext}$ ($\tilde\omega_{\rm ext}=1/262$ corresponds to the experimental situation). 
Remarkably, the curve that gives the renormalization of the dipole frequency appears to be independent of the external potential strength in the limit $\omega_{\rm ext} \ll h/(md^2)$ (see \cite{supp}). This property differs from the low depth result \cite{inguscio2001b,science2001,stringari2002}. Additionally, the dipole mode that we excite turns out to be independent of the strength of the interaction. This well-known feature for a single well remains valid for our chain of coupled wells. This was also pointed out in \cite{stringari2002} for large amplitude oscillations in the low depth limit.  %In the micromotion that we consider, there is no atoms on the other side of the barrier at the first turning point (point $C$ in Fig.~\ref{fig1}). This suppresses the possibility of a modification of the dynamics by the interactions. It is in contrast with the studies about Josephson oscillations for which interactions dominate the dynamics \cite{Josephson}. 
We have numerically checked these two properties, and we can therefore precisely infer the potential depth from the first dipole oscillation period. We experimentally measure a period $T= 106\:\pm$ 4 $\mu$s  ($\tilde \omega T=4.36\pm0.14$), that corresponds to $s_0=3.21 \pm 0.12$ according to Fig.~\ref{fig2}a (arrow). For this depth, the potential accommodates for two bound states \cite{supp}.

Another consequence of the collective nature of the dipole oscillation is the quasi-independence of the oscillation period on the initial angle $\theta_0$ (see \cite{supp}). This is illustrated in Fig.~\ref{fig2}b where we compare for different initial angle $\theta_0$ the period measured experimentally, the result of the numerical simulation in the same condition as in the experiment and with the calibration explained above, and the classical prediction. Within the error bars, the oscillation period of the dipole mode is constant in the wide range of angle that we studied ($30\leq \theta_0 \leq 80^o$), an observation confirmed by numerical simulations (see \cite{supp}). This is to be contrasted with the classical prediction where the non-harmonicities of the wells yield an oscillation period that strongly depends on the initial angle. Furthermore, the BEC being put in a far out-of-equilibrium state by the initial sudden displacement, we numerically checked that its relaxation towards equilibrium does not modify the oscillation dynamics studied above: the relaxation occurs on a timescale of a few ms, thus much larger than the oscillation period. 

\begin{figure}[h!]
\centering
\includegraphics[width=8cm]{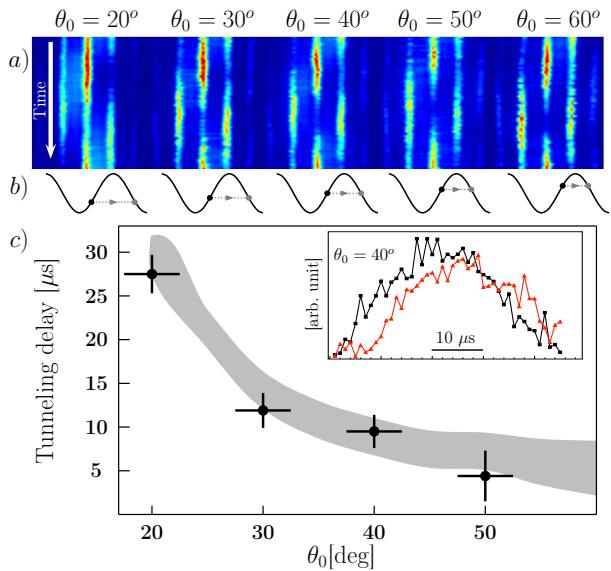}
\caption{(a) Experimental images: zoom on $C$, $D_1$ and $D_2$ packets for various values of the initial offset angle of the lattice $\theta_0$.
(b) Sketch of the tunneling process for different $\theta_0$ showing the thickness of the potential barrier at the first turning point (point $C$).
(c) Quantitative tunneling time delay extracted from the experimental images (black disk), compared to the results of the numerical simulations without adjustable parameter but including the uncertainty on the lattice depth (gray area). Inset: Integrated density distribution of the reflected $D_1$ (black square) and tunneled $D_2$ (red triangles) packets for $\theta_0=40^o$.}
\label{fig3}
\end{figure}

Let us analyze further the experimental time sequence of Fig.~\ref{fig1}.
We observe  that the centers of the packets $D_1$ and $D_2$ do not coincide in time. The packet $D_2$ that has tunneled through the potential barrier is delayed with respect to $D_1$. The independence of the period on the initial angle $\theta_0$ ensures that the time delay that we measure is not affected by anharmonicities. The analysis in momentum space facilitates the precise measurement of the time delay since the two packets that split on the barrier acquire an opposite momentum after a quarter of period. We investigated systematically this delay through a set of experiments, with 1 $\mu$s time step, for which the initial angle, and thus the initial sudden displacement of the lattice, is varied. The data are presented in Fig.~\ref{fig3}a. We observe that when increasing $\theta_0$, the relative delay between the reflected ($D_1$) and tunneled ($D_2$) wave packets decreases. Indeed, as intuitively expected from semiclassical approach, for a larger shift angle, the thickness of the potential barrier at the turning point decreases and the turning point corresponds to a higher energy (see Fig.~\ref{fig3}b). To extract the time delay between the wave packets, we determine the maximum of the density distributions of the $D_1$ and $D_2$ wave packets (see Inset of Fig.~\ref{fig3}) using as a guide a Gaussian fit of the density distributions. For $\theta_0 \ge 60^o$, the delay cannot be measured precisely because of the small thickness of the barrier which yields a partial overlap of packets $B$ and $D_2$.  The numerical simulations are in very good agreement with the experimental data without any adjustable parameter once the calibration has been carefully performed (see Fig.~\ref{fig3}c). The error bars on the numerical simulations are due to the uncertainty on the experimentally measured period. Remarkably, the numerical simulations show that interactions do not affect the value of the delay time. For the smallest displacements ($\theta_0 = 20^o$ and $\theta_0 = 30^o$), the dynamics is completely dominated by the bound Bloch states and the time delay that we measure is therefore exclusively due to the tunnel effect (see \cite{supp}). The contribution of the unbound states grows with $\theta_0$ to reach 35\% for $\theta_0 = 50^o$.

In the last part of the time evolution represented in Fig.~\ref{fig1}, we observe that the packet $F$ has a momentum opposed to that of packet $B$. Such behavior may appear surprising at first sight: indeed in the absence of tunneling (classical prediction), one would expect the packet $F$ to have exactly the opposite momentum compared to what we observe. This counterintuitive behavior results from a well-known interference effect occurring in a Mach-Zehnder interferometer (MZI) with symmetric beam paths and 50-50 beam splitters: the exit channel is the opposite to the entrance channel. The packet of atoms in $B$ has positive momentum components centered about $h/d$. We define this state as $|B\rangle=|p_0 \rangle$. The potential barrier at the turning point $C$ plays the role of a coherent beam splitter: $|D \rangle=\cos \varphi |p_0 \rangle + i\sin \varphi |-p_0 \rangle$. The confinement driving the subsequent evolution plays the role of mirrors. At the turning point $E$ the two packets that have been split are recombined on a second beam splitter. The result of this recombination is read out in momentum space in $F$: $|F\rangle=((\cos \varphi |p_0 \rangle + i\sin \varphi|-p_0 \rangle)\cos \varphi + i\sin \varphi(\cos \varphi |-p_0 \rangle + i\sin \varphi|p_0 \rangle)=\cos 2\varphi |p_0 \rangle + i\sin 2\varphi |-p_0 \rangle$. We infer that the momentum in $F$ is the exact opposite to that in $B$ for a perfect 50-50 beam splitter ($\varphi=\pi/4$). %This result is well-known in optics for identical path MZI: the exit channel is the opposite to the entrance channel. 

In our lattice configuration, the interfering packets originate from two adjacent sites (see Fig.~\ref{fig3}b). We therefore have multiple MZI working simultaneously and addressing only the external degree of freedom. We analyze the exit of the interferometers by interference in momentum space. The constructive interference observed in Fig.~\ref{fig1} is thus due to the persistence of the global coherence of the BEC on the dipole micromotion period timescale. We numerically checked that this interference effect is immune to interactions. It however depends on the lattice depth, which changes the ratio between the tunneled and the reflected parts of the wave packet, and therefore the population that ends up with the same momentum as in the entrance channel $B$.
The constructive interference observed in Fig.~\ref{fig1} corresponds to a ratio  $\Pi_{-1}/(\Pi_1+\Pi_{-1})=$91 \% where $\Pi_n$ is the number of atoms with momentum $nh/d$ (corresponding to $\varphi\simeq \pi/5$). 
Finally, after an extra quarter of period, the atoms from $F$ are back in the initial state $A$. Therefore, when increasing the evolution time in the shifted lattice, the dynamics shown in Fig.~\ref{fig1} repeats again and the sensitivity of the coupled chain of MZI interferometers increases. 

In conclusion, we have performed a direct measurement of the tunneling delay time through the barriers of an optical lattice by studying the time evolution of a BEC after a sudden displacement of the lattice. The excitation of the BEC by a displacement of a few tens of nm provides a robust method to calibrate the lattice depth independently of the external potential, the interactions and the value of the displacement. As a perspective, this experiment offers the possibility of controlling the tunneling delay time with time-dependent barriers in close analogy with the Landauer problem \cite{landauer}. Additionnally we have observed the constructive interference in a MZI provided by two successive interactions with the barriers acting as beam splitter. Such a micron size interferometer could be of interest to measure locally short-range force at the vicinity of a surface \cite{c12,c13,c14,c15,c16}.

We are indebted to S. Faure and L. Polizzi for their precious technical assistance for the building up of our new experimental setup.
We thank J. Dalibard, J. G. Muga, B. Georgeot and J. Vigu\'e for useful comments. We are grateful to C. Besse and R. Duboscq for the numerous fruitful discussions we had on the use of the numerical GPELab toolbox. This work was supported by Programme Investissements d'Avenir under the program ANR-11-IDEX-0002-02, reference ANR-10-LABX-0037-NEXT.


\begin{thebibliography}{99}
\bibitem{classical}
According to classical mechanics, the oscillation period $T_{\rm c}$ in the lattice wells depends on the initial shift angle $\theta_0$: $T_{\rm c}(\theta_0)=(2md^2/h)s^{-1/2}f(\theta_0)$ with $f(\theta_0)=(2/\pi)\int_0^{\theta_0} (\cos^2 \theta-\cos^2 \theta_0)^{-1/2} d\theta$ since there is no initial velocity. $T_c$ is marginally affected by the external confinement since $\omega_{\rm ext} \ll h/(md^2)$.
\end{thebibliography}
\end{document}